\documentclass[trackchanges,twocolumn]{aastex7}

\begin{document}

\title{Population-Dependent $r$-process Scatter in the Globular Cluster M15}

\author[0000-0002-0786-7307]{Lauren E. Henderson}
\affiliation{Department of Physics and Astronomy, University of Notre Dame,
225 Nieuwland Science Hall,
Notre Dame, IN 46556, USA}
\email[show]{lhender6@nd.edu}  

\author[0000-0003-0398-639X]{Roman Gerasimov}
\affiliation{Department of Physics and Astronomy, University of Notre Dame,
225 Nieuwland Science Hall,
Notre Dame, IN 46556, USA}
\email{rgerasim@nd.edu}

\author[0000-0001-6196-5162]{Evan N. Kirby}
\affiliation{Department of Physics and Astronomy, University of Notre Dame,
225 Nieuwland Science Hall,
Notre Dame, IN 46556, USA}
\email{ekirby@nd.edu}

\begin{abstract}
Multiple populations, defined by correlations between light element abundances, are an almost ubiquitous property of globular clusters. On the other hand, dispersions among the heavy elements are limited to a few rare clusters. In this letter, we present Mg, Y, Ba, La, and Eu measurements for 89 stars in M15 with errors $<$ 0.4 dex from Keck/DEIMOS medium-resolution spectra. We find higher Ba, La, and Eu dispersions in the first generation of stars than in the second generation at a significance of $\ge$2 $\sigma$. This is evidence for inhomogeneous mixing of gas during the formation of the first generation of stars, which subsequently became well-mixed prior to the formation of the second generation of stars. If the $r$-process event that caused the abundance dispersions was born with the first population of stars, it must be an $r$-process site with a short delay time.
\end{abstract}



\section{Introduction} \label{sec:intro}


Globular clusters (GCs) are not simple stellar populations. Contrary to what would be assumed if all the stars formed simultaneously with little time for chemical evolution, GCs exhibit significant spread in the abundances of light elements \citep[e.g.,][]{freeman_omegacen_1975, meszaros_lightelements_2015, bastian_lardo_MPs_2018, milone_clusters_2022}. These variations in abundances among stars in a GC define multiple populations within the cluster. In fact, most GCs show evidence of multiple populations. This can be seen spectroscopically by directly measuring abundances \citep[e.g.,][]{peterson_M13_1980,sneden_M15_1997, carretta_NaO_2009, marino_NaO_2011, masseron_apogee_2019}, as well as photometrically because the differences in chemical abundances cause a spread in the color-magnitude diagram of the cluster \citep[e.g.,][]{bedin_OmegaCen_2004, smolinski_sdss_2011, milone_hubbleGC_2017, gerasimov_47Tuc_2024}.

The abundances of light elements in GCs exhibit correlations and anticorrelations that are hallmarks of multiple populations. These include the Na--O anticorrelation, the Al--Mg anticorrelation, and the N--C anticorrelation. Stars that resemble field stars are deemed to be in Population 1 (now also known as the first generation), while those that have enhanced N, Na, and Al and depleted C, O, and Mg are in Population 2 (second generation) and are not commonly found outside of GCs. There is significant variation between clusters as to the range of abundances and the number of populations \citep{gratton_review_2019}, but these abundance features are present in most old GCs. There is still no consensus as to the cause of these anticorrelations that is consistent with all GC properties \citep[size, relative size of the populations, age, abundances, etc.,][]{bastian_lardo_MPs_2018}. 

While the variations in light element abundances are distinctive of GCs, 
they exhibit very low dispersions of iron and iron-peak elements \citep{willman_galaxydef_2012}. Thus, it has long been assumed that they should show little to no variation in the abundances of elements past the iron peak. 

Some metal-poor GCs undermine this assumption. Multiple low-metallicity clusters, such as M15 and M92 ($\rm[Fe/H]=-2.37$ and $\rm[Fe/H]=-2.31$, respectively \citealt{harris_GCs_1996, harris_GCs_2010}), display dispersions in neutron-capture element abundances \citep{sneden_M15_1997, roederer_m92_2011, roederer_metalpoor_2011, sobeck_m15_2011, kirby_r-process_2023, cabreragarcia_abundances_2024, schiappacasse-ulloa_GCn-capture_2024}. Based on the old ages of these clusters and their [Ba/Eu] ratios, the majority of the neutron-capture enrichment must be due to the $r$-process, as intermediate-mass asymptotic giant branch stars would not evolve rapidly enough to supply significant $s$-process enrichment \citep{sneden_M15_2000}.

A dispersion in $r$-process enrichment could either be a property the stars form with, or it could be due to an $r$-process event near the cluster that preferentially enriched some stars over others. In the case of M15, the latter explanation was ruled out by \citet{kirby_m15_2020}. This means that there are GCs that form with a spread in neutron-capture abundances.

Further clues regarding heavy element dispersions in GCs came with the study of M92 by \citet{kirby_r-process_2023}. They found that the stars in Population 1, the first generation, have a higher dispersion in $r$-process abundances than the second generation stars. They proposed that this discrepancy in $r$-process dispersion between populations suggests the populations formed at different times. The first generation was enriched with the $r$-process during its formation, giving some stars much higher $r$-process abundances than others. By the time the second generation formed, the $r$-process material was well-mixed, resulting in a low $r$-process dispersion.

A larger $r$-process dispersion in the first generation than in the second generation has since been observed in NGC 2298 as well \citep{bandyopadhyay_ngc2298_2025}. Whether this is ubiquitous among GCs, applicable to only low-metallicity GCs, or unique to M92 and NGC 2298 is an open question. Searching for this abundance pattern in other metal-poor GCs with known $r$-process dispersions, such as M15, is a place to begin answering this question. \citet{schiappacasse-ulloa_GCn-capture_2025} looked for population- dependent neutron-capture abundance scatter in M15 (among other GCs), but were limited by a small sample size. Medium-resolution spectroscopy, which has been shown to be sufficient for measuring neutron-capture abundances \citep{duggan_neutron_2018, kirby_m15_2020, henderson_491_2025}, offers the opportunity to measure the neutron-capture abundances in many stars in M15 to make a detailed comparison between populations.

In this Letter, we present Mg, Y, Ba, La, and Eu abundances  with errors $<$ 0.4 dex in 89 stars in the GC M15 from medium-resolution Keck/DEIMOS spectra. The Mg measurements allow us to compare the neutron-capture element abundances for each population in M15.

In Section \ref{sec:data}, we describe the data used. We discuss the abundance measurements in Section \ref{sec:abunds}. In Section \ref{sec:results}, we discuss the abundance results, including the dispersions of the $r$-process elements in each population. Finally, we conclude with the implications for multiple populations in GCs and the $r$-process in Section \ref{sec:discussion}.

\section{Data} \label{sec:data}

We used existing observations of M15 obtained with the DEIMOS spectrograph \citep{faber_deimos_2003} on the Keck~II telescope.  Two slitmasks were observed with the 900ZD grating \citep{kirby_carbon_2015} and one slitmask was observed with the 1200B grating \citep{kirby_m15_2020}.  The 900ZD exposures totaled 30 and 45 minutes for the two respective slitmasks, and the 1200B exposures totaled 4.3~hours.

The spectral range of 900ZD spectra is approximately 4000--7400~\AA\ with a resolution of $\Delta \lambda = 2.0$~\AA\ FWHM\@.  The spectral range of 1200B is approximately 3900--6500~\AA\ with a resolution of $\Delta \lambda = 1.1$~\AA\ FWHM\@.

All three slitmasks additionally have redder spectral coverage with the 1200G grating \citep{kirby_lithium-rich_2016}.  This configuration provides a spectral range of approximately 6300--9100~\AA\ with $\Delta \lambda = 1.2$~\AA\@.

M15 is significantly more metal-poor ($\mathrm{[Fe/H]}<-2$) than the majority of foreground field stars. The measured metallicity distribution of stars (see Sec.~\ref{sec:abunds}) is distinctly bimodal with peaks near $\mathrm{[Fe/H]}\sim-2.5$ ($\sim90\%$ of the stars) and $\mathrm{[Fe/H]}\sim-0.3$ ($\sim10\%$ of the stars). The two modes of the distribution are well-isolated, since we found no stars with $-1.94<[\mathrm{Fe/H}]<-0.78$. The distinct gap allows metallicity-based membership selection without biasing the sample. We rejected all stars with $\mathrm{[Fe/H]}>-1$ as non-members.

Further rejection of non-members was carried out by cross-matching the coordinates of the observed stars with the Gaia DR3 catalog \citep{GaiaDR3_2023}. We rejected stars with proper motions deviating from the cluster median ($(\mu_\alpha \cos(\delta),\mu_\delta)=(-0.7,-3.9)\ \mathrm{mas}\ \mathrm{yr}^{-1}$) by over $2\ \mathrm{mas}\ \mathrm{yr}^{-1}$ in any direction (the proper motion dispersion in M15 is $\sim 0.3\ \mathrm{mas}\ \mathrm{yr}^{-1}$). We also rejected stars with multiple Gaia cross-matches within $1\ \mathrm{arcsec}$ if their $G$-magnitudes differ by less than $1\ \mathrm{mag}$ (otherwise the brightest match was taken).

\begin{figure*}
    \centering
    \includegraphics[width=1\linewidth]{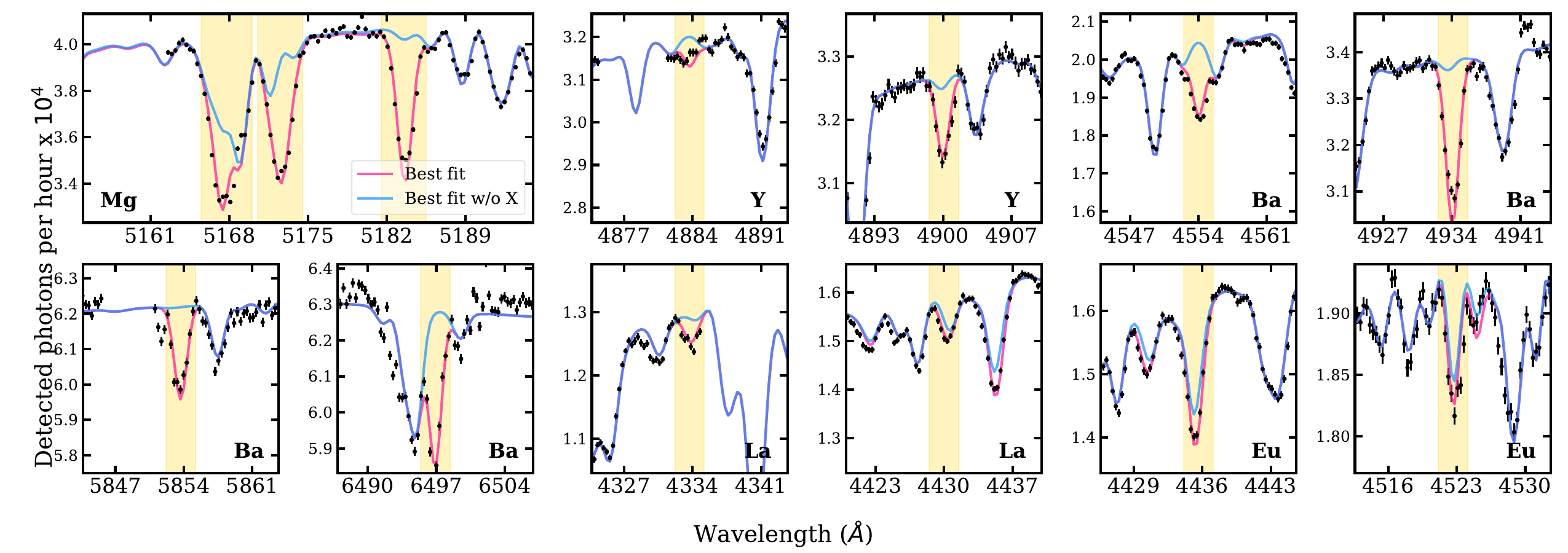}
    \caption{Select Mg and neutron-capture element absorption lines for the star 55914 in M15. The relevant lines in each panel are highlighted in yellow. The pink spectrum is the best fit model spectrum. The blue model spectrum has the same best fit stellar parameters but arbitrarily low abundances of the relevant elements.}
    \label{fig:spectrum}
\end{figure*}

\begin{deluxetable}{cccccccccc}
\tablecaption{Multi-Element Abundance Catalog of M15 Stars \label{tab:abundances}}
\tablehead{\colhead{Name} & \colhead{S/N blue} & \colhead{Temp (K)} & \colhead{$\log g$} & \colhead{[Fe/H]} & \colhead{[Mg/Fe]} & \colhead{[Y/Fe]} & \colhead{[Ba/Fe]} & \colhead{[La/Fe]} & \colhead{[Eu/Fe]}}
\startdata
51057 & 50.49 & 4707 & 1.47 & -2.65$\pm$0.02 & $+0.58^{+0.01}_{-0.01}$ & $-0.20^{+0.03}_{-0.04}$ & $+0.16^{+0.02}_{-0.02}$ & $+0.01^{+0.04}_{-0.07}$ & $+0.40^{+0.03}_{-0.22}$ \\[2mm]
50051 & 50.33 & 4996 & 1.65 & -2.68$\pm$0.04 & $+0.71^{+0.03}_{-0.03}$ & $-0.09^{+0.09}_{-0.14}$ & $+0.32^{+0.08}_{-0.06}$ & \nodata & $+0.52^{+0.15}_{-0.50}$ \\[2mm]
38026 & 60.85 & 4436 & 0.63 & -2.44$\pm$0.02 & $+0.65^{+0.01}_{-0.01}$ & $-0.18^{+0.03}_{-0.05}$ & $+0.25^{+0.02}_{-0.04}$ & $+0.30^{+0.03}_{-0.03}$ & $+0.58^{+0.03}_{-0.09}$ \\[2mm]
49658 & 74.51 & 4611 & 1.26 & -2.56$\pm$0.02 & $+0.58^{+0.01}_{-0.01}$ & $-0.10^{+0.03}_{-0.10}$ & $+0.30^{+0.03}_{-0.04}$ & $+0.53^{+0.05}_{-0.05}$ & $+0.78^{+0.09}_{-0.25}$ \\[2mm]
60808 & 20.27 & 4546 & 1.20 & -2.54$\pm$0.01 & $+0.51^{+0.01}_{-0.02}$ & $-0.32^{+0.08}_{-0.02}$ & $+0.31^{+0.01}_{-0.18}$ & $+0.34^{+0.02}_{-0.04}$ & $+0.87^{+0.05}_{-0.05}$ \\[2mm]
66116 & 43.55 & 6106 & 2.83 & -2.45$\pm$0.08 & $+0.80^{+0.04}_{-0.06}$ & $<+0.34$ & $+0.34^{+0.09}_{-0.15}$ & $+1.55^{+0.11}_{-0.22}$ & $<+1.76$ \\[2mm]
61191 & 36.46 & 5359 & 2.87 & -2.40$\pm$0.05 & $+0.42^{+0.02}_{-0.04}$ & $+0.22^{+0.09}_{-0.18}$ & $+0.54^{+0.09}_{-0.08}$ & $+1.17^{+0.09}_{-0.08}$ & $+1.16^{+0.13}_{-0.48}$ \\[2mm]
16135 & 78.27 & 4695 & 0.47 & -2.68$\pm$0.01 & $-0.12^{+0.08}_{-0.01}$ & $-0.14^{+0.14}_{-0.01}$ & $+0.34^{+0.01}_{-0.03}$ & $+0.18^{+0.09}_{-0.02}$ & $+0.69^{+0.04}_{-0.05}$ \\[2mm]
7815 & 59.94 & 6217 & 3.15 & -2.81$\pm$0.21 & $+0.68^{+0.07}_{-0.18}$ & $<+0.96$ & $+0.02^{+0.28}_{-0.44}$ & \nodata & \nodata \\[2mm]
49428 & 110.85 & 4828 & 1.60 & -2.63$\pm$0.02 & $+0.20^{+0.02}_{-0.02}$ & $-0.06^{+0.05}_{-0.03}$ & $+0.37^{+0.02}_{-0.04}$ & $+0.36^{+0.05}_{-0.04}$ & $+0.66^{+0.05}_{-0.10}$ \\[2mm]
11998 & 88.34 & 5486 & 3.39 & -2.51$\pm$0.06 & $+0.47^{+0.02}_{-0.04}$ & $<+0.15$ & $+0.01^{+0.06}_{-0.16}$ & \nodata & \nodata \\[2mm]
49479 & 52.15 & 4697 & 1.37 & -2.57$\pm$0.02 & $+0.61^{+0.02}_{-0.01}$ & $-0.37^{+0.08}_{-0.12}$ & $-0.02^{+0.06}_{-0.05}$ & $<-0.14$ & $+0.38^{+0.11}_{-0.21}$ \\[2mm]
57577 & 50.29 & 4797 & 1.63 & -2.52$\pm$0.03 & $+0.63^{+0.02}_{-0.02}$ & $+0.12^{+0.05}_{-0.11}$ & $+0.42^{+0.04}_{-0.08}$ & $+0.60^{+0.04}_{-0.09}$ & $+1.08^{+0.10}_{-0.09}$ \\[2mm]
30961 & 79.20 & 4339 & 0.56 & -2.49$\pm$0.03 & $+0.51^{+0.02}_{-0.01}$ & $-0.39^{+0.10}_{-0.02}$ & $-0.21^{+0.04}_{-0.02}$ & $+0.12^{+0.04}_{-0.04}$ & $+0.58^{+0.03}_{-0.06}$ \\[2mm]
56734 & 66.14 & 4639 & 1.28 & -2.52$\pm$0.02 & $+0.58^{+0.02}_{-0.01}$ & $-0.19^{+0.06}_{-0.06}$ & $+0.31^{+0.03}_{-0.05}$ & $+0.40^{+0.03}_{-0.07}$ & $+0.69^{+0.09}_{-0.17}$ \\[2mm]
55914 & 87.29 & 4664 & 1.32 & -2.60$\pm$0.01 & $+0.53^{+0.01}_{-0.01}$ & $-0.13^{+0.07}_{-0.04}$ & $+0.41^{+0.01}_{-0.03}$ & $+0.53^{+0.01}_{-0.12}$ & $+0.91^{+0.01}_{-0.36}$ 
\enddata
\tablecomments{Stellar parameters and abundances for a subset of M15 stars in this study. A full version of this table, including coordinates and upper limits, is available in a machine-readbale format. S/N blue is the signal-to-noise of the blue spectrum (900ZD or 1200B), where most of the neutron-capture absorption lines are located.}
\end{deluxetable}

\section{Abundance Measurements} \label{sec:abunds}

The stellar parameters and element abundances were determined by fitting forward spectral models to the observed spectra. The approach adopted in this work is broadly similar to that of \citet{fisher_chemfit_2024} with multiple major improvements that are briefly introduced below. A more complete discussion of our spectral models and fitting routines is deferred to a future publication (Gerasimov et al. 2025, \textit{in preparation}).

The spectral models employed in our analysis were calculated using the \texttt{ATLAS-9}/\texttt{SYNTHE}/\texttt{BasicATLAS} setup \citep{ATLAS5,ATLAS9_1,ATLAS9_2,SYNTHE,BasicATLAS}. The atomic line list was compiled by combining the lists in \citet{Kurucz_atoms,kirby_grids_2011,VALD3_atoms,escala_elemental_2019} and by manually adjusting the oscillator strengths of selected lines, in order to obtain the best correspondence with the high-resolution spectrum of Arcturus \citep{arcturus_spectrum}. We further included the $\mathrm{CH}$ molecular lines from \citet{CH_masseron}, $\mathrm{MgH}$ lines from \citet{MgH_masseron}, and the ``standard set'' of other diatomic molecules from \citet{kurucz_lines}. However, we did not include $\mathrm{TiO}$ and $\mathrm{H_2O}$ lines in our models to improve computational performance. The effect of these molecules on the spectrum is expected to be very small at the effective temperatures considered in this work ($T_\mathrm{eff}>4000\ \mathrm{K}$).

Our analysis was carried out in three stages. At each stage, the same model was simultaneously fit to all available spectra of the star, as well as the Gaia DR3 \citep{GaiaDR3_2023} $\mathrm{BP}-\mathrm{RP}$ color of the star. The latter was accomplished by calculating synthetic photometry of the spectral model using the \texttt{synphot} utility of \texttt{BasicATLAS}, and adopting the optical reddening of $E(B-V)=0.1$ \citep{harris_GCs_1996}, the total-to-selective extinction of $R_V=3.1$ and the reddening law from \citet{extinction}. The addition of broadband photometry to the fit alleviates the degeneracy between temperature and metallicity by providing additional information about the overall spectral energy distribution of the star.

At the first stage, the spectral models were generated by linearly interpolating a pre-computed regular 5D grid ($T_\mathrm{eff}$, $\log(g)$, $[\mathrm{M/H}]$, $[\mathrm{\alpha/M}]$ and $[\mathrm{C/M}]$, where $\mathrm{M}$ refers to all metals, and $\mathrm{\alpha}$ to $\mathrm{O}$, $\mathrm{Ne}$, $\mathrm{Mg}$, $\mathrm{Si}$, $\mathrm{S}$, $\mathrm{Ar}$, $\mathrm{Ca}$ and $\mathrm{Ti}$). The best-fit model parameters as well as radial velocity and spectral resolution were obtained using the implementation of the \textit{Trust Region Reflective} (TRR) algorithm in \texttt{scipy} \citep{scipy}. The best-fit radial velocity and spectral resolution obtained at this stage were fixed and reused in the remaining stages of the analysis.

At the second stage, the models were generated on demand during the fitting process (hereinafter referred to as the \textit{live synthesis mode}), and fitted to the observations using TRR\@. We allowed for 7 degrees of freedom: $T_\mathrm{eff}$, $\log(g)$, $[\mathrm{M/H}]$, $[\mathrm{\alpha'/M}]$, $[\mathrm{C/M}]$, $[\mathrm{O/M}]$ and $[\mathrm{Fe/M}]$. $\mathrm{\alpha'}$ refers to the same set of elements as $\mathrm{\alpha}$ in the first stage with the exception of oxygen, as its abundance was included as a distinct degree of freedom ($[\mathrm{O/M}]$). $[\mathrm{Fe/M}]$ refers to the abundance of iron compared to the overall metallicity offset, such that $[\mathrm{Fe/H}]=[\mathrm{M/H}]+[\mathrm{Fe/M}]$. Live synthesis fitting is, by far, the most computationally demanding stage of our analysis, which is made possible by the small number of stars considered in this work ($\sim 100$), and our exclusion of $\mathrm{TiO}$ and $\mathrm{H_2O}$ opacities from the models. This stage is necessary because GCs display large member-to-member variations in carbon and oxygen abundances due to both multiple populations \citep{bastian_lardo_MPs_2018} and evolutionary processing \citep{thermohaline_mixing,first_dredge_up}. If unaccounted for, $[\mathrm{C/M}]$ and $[\mathrm{O/M}]$ have a large impact on the entire spectrum due to their effect on electron density and continuum opacity in the atmosphere. By allowing both $[\mathrm{C/M}]$ and $[\mathrm{O/M}]$ to vary during the fitting process, we are able to isolate genuine member-to-member variations in $r$-process abundances from systematic knock-on effects of light elements. We note that measuring carbon and oxygen abundances individually is challenging due to their involvement in molecular chemistry alongside nitrogen. For this reason, we treat $[\mathrm{C/M}]$ and $[\mathrm{O/M}]$ as ``nuisance parameters'', i.e. their inclusion is expected to improve the accuracy of other best-fit parameters but may not necessarily provide reliable constraints on the carbon and oxygen abundances themselves.

At the third stage of the analysis, we fixed all of the parameters determined in the previous two stages, and varied the abundances of $18$ individual elements: $[\mathrm{Li/M}]$, $[\mathrm{Na/M}]$, $[\mathrm{Mg/\alpha'}]$, $[\mathrm{Al/M}]$, $[\mathrm{Si/\alpha'}]$, $[\mathrm{K/M}]$, $[\mathrm{Ca/\alpha'}]$, $[\mathrm{Sc/M}]$, $[\mathrm{Ti/\alpha'}]$, $[\mathrm{V/M}]$, $[\mathrm{Cr/M}]$, $[\mathrm{Mn/M}]$, $[\mathrm{Co/M}]$, $[\mathrm{Ni/M}]$, $[\mathrm{Y/M}]$, $[\mathrm{Ba/M}]$, $[\mathrm{La/M}]$ and $[\mathrm{Eu/M}]$\@. The model spectra were generated under the assumption that the effects of individual elements on the 2D (wavelength and optical depth) atmospheric opacity are additive and independent of each other. At each iteration of the fitting process, we evaluated the \textit{response functions} (i.e., offsets in 2D opacity) due to each element separately, summed the response functions to determine the complete 2D opacity of the atmosphere and solved the equation of radiative transfer (using \texttt{SYNTHE}, \citealt{SYNTHE}) to obtain the model spectrum. When calculating the response functions of most elements, we assumed that the effect of each element is restricted to the lines of that element, as well as its ions and molecules. However, $[\mathrm{Na/M}]$, $[\mathrm{Mg/\alpha'}]$, $[\mathrm{Al/M}]$, $[\mathrm{Si/\alpha'}]$, and $[\mathrm{Ca/\alpha'}]$ also noticeably impact the ionization balance of the atmosphere and, therefore, the line strengths of other elements as well as continuum opacity. When calculating the response functions of these five elements, we included the offsets due to all line opacities, as well as continuum opacities in the vicinity of those lines.

The best-fit parameters at the third stage were determined using the Goodman--Weare Markov chain Monte Carlo (MCMC) method \citep{Goodman_Weare_MCMC} implemented in the \texttt{emcee} package \citep{emcee_2013}. Each chain employed $36$ walkers and $5000$ steps per walker, of which the first $300$ were discarded as burn-in.
%
%
%
%
We chose to explore the entire likelihood space with MCMC instead of merely fitting the model with an optimizer (TRR) in order to capture the asymmetric posteriors and differentiate between measurements and upper limits. The upper and lower asymmetric error bars are the 16th and 84th percentiles, respectively. We used these percentile values to determine upper limits. If the width of the MCMC posterior distribution between the 16th and 84th percentiles is greater than 1~dex and the distribution is skewed such that the tail extends to lower abundances, the abundance is best described as an upper limit. These are flagged in Table \ref{tab:abundances} as upper limits. Figure~\ref{fig:spectrum} shows portions of an observed spectrum, along with the best-fit spectral model.

We compared the abundances of matching stars with high-resolution measurements from \citet{sneden_BaNa_2000}, \citet{worley_M15_2013}, \citet{sobeck_m15_2011}, \citet{meszaros_apogee_2020}, and \citet{cabreragarcia_abundances_2024}. The average absolute offset between our neutron-capture abundances and high-resolution measurements is 0.18 dex. Our Mg abundances are $\approx0.2$ dex higher than those in \citet{meszaros_apogee_2020} but have a similar spread. Offsets from high-resolution measurements are not a concern in this study, as they do not trend with [Mg/Fe] and thus have no dependence on population.

\begin{deluxetable*}{ccccccc}[t]
\tablecolumns{7}
\tablewidth{0pt}
\tablecaption{MCMC Results\label{tab:mcmcresults}}
\tablehead{
 \colhead{Element} & \colhead{Error cut} & \colhead{1G $\rm\overline{[X/Fe]}$} & \colhead{2G $\rm\overline{[X/Fe]}$} & \colhead{1G $\rm\sigma_{[X/Fe]}$}  & \colhead{2G $\rm\sigma_{[X/Fe]}$} & \colhead{Notes}}
\startdata
Y & 0.1 & $-0.21^{+0.04}_{-0.04}$ & $-0.19^{+0.04}_{-0.04}$ & $0.21^{+0.04}_{-0.04}$ & $0.13^{+0.05}_{-0.03}$ & \\[2mm]
Y & 0.4 & $-0.16^{+0.03}_{-0.03}$ & $-0.19^{+0.05}_{-0.05}$ & $0.23^{+0.03}_{-0.02}$ & $0.19^{+0.05}_{-0.04}$ & \\[2mm]
\hline
Ba & 0.1 & $0.21^{+0.04}_{-0.04}$ & $0.32^{+0.03}_{-0.03}$ & $0.26^{+0.03}_{-0.03}$ & $0.13^{+0.03}_{-0.02}$ & \\[2mm]
Ba & 0.4 & $0.22^{+0.03}_{-0.03}$ & $0.33^{+0.03}_{-0.03}$ & $0.26^{+0.03}_{-0.02}$ & $0.14^{+0.03}_{-0.02}$ & \\[2mm]
\hline
La & 0.1 & $0.44^{+0.06}_{-0.06}$ & $0.38^{+0.04}_{-0.04}$ & $0.33^{+0.06}_{-0.04}$ & $0.15^{+0.04}_{-0.03}$ & \\[2mm]
La & 0.4 & $0.45^{+0.05}_{-0.05}$ & $0.42^{+0.04}_{-0.04}$ & $0.35^{+0.04}_{-0.04}$ & $0.16^{+0.04}_{-0.03}$ & \\[2mm]
\hline
Eu & 0.1 & $0.76^{+0.09}_{-0.09}$ & $0.65^{+0.08}_{-0.08}$ & $0.36^{+0.09}_{-0.07}$ & $0.26^{+0.10}_{-0.06}$ & \\[2mm]
Eu & 0.1 & $0.76^{+0.10}_{-0.09}$ & $0.72^{+0.05}_{-0.05}$ & $0.36^{+0.09}_{-0.06}$ & $0.12^{+0.06}_{-0.04}$ & 54055 excluded\\[2mm]
Eu & 0.4 & $0.70^{+0.05}_{-0.06}$ & $0.66^{+0.06}_{-0.06}$ & $0.30^{+0.04}_{-0.04}$ & $0.22^{+0.06}_{-0.04}$ & \\[2mm]
Eu & 0.4 & $0.70^{+0.05}_{-0.05}$ & $0.71^{+0.04}_{-0.04}$ & $0.30^{+0.04}_{-0.04}$ & $0.12^{+0.05}_{-0.04}$ & 54055 excluded\\
\enddata
\end{deluxetable*}

\section{Results} \label{sec:results}

\begin{figure}
    \centering
    \includegraphics[width=1\linewidth]{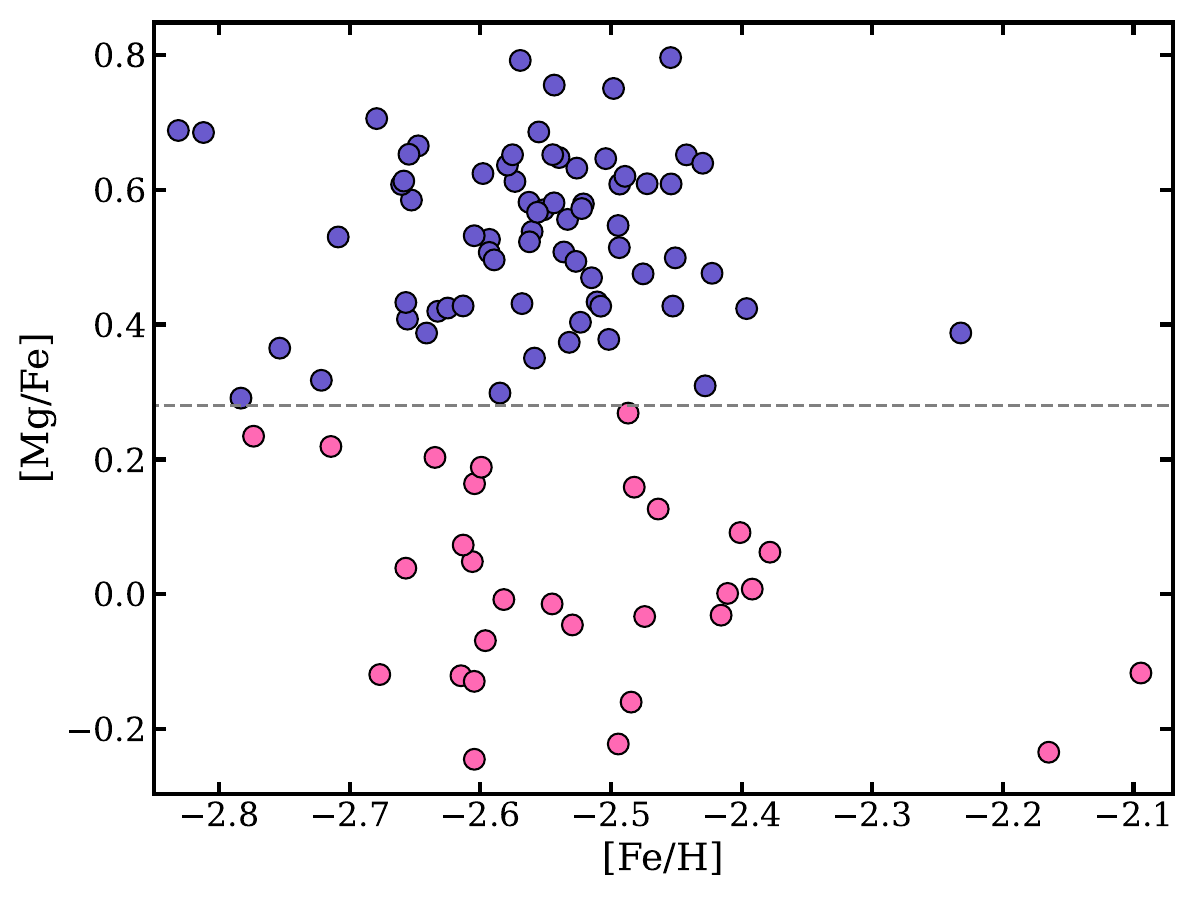}
    \caption{[Mg/Fe] v.\ [Fe/H] for stars in M15. The points are colored according to their the k-means clustering. The dashed gray line marks [Mg/Fe] = 0.28, which we use to separate the stars into populations for calculating mean and dispersion.}
    \label{fig:Kmeans}
\end{figure}

\begin{figure}
    \centering
    \includegraphics[width=1\linewidth]{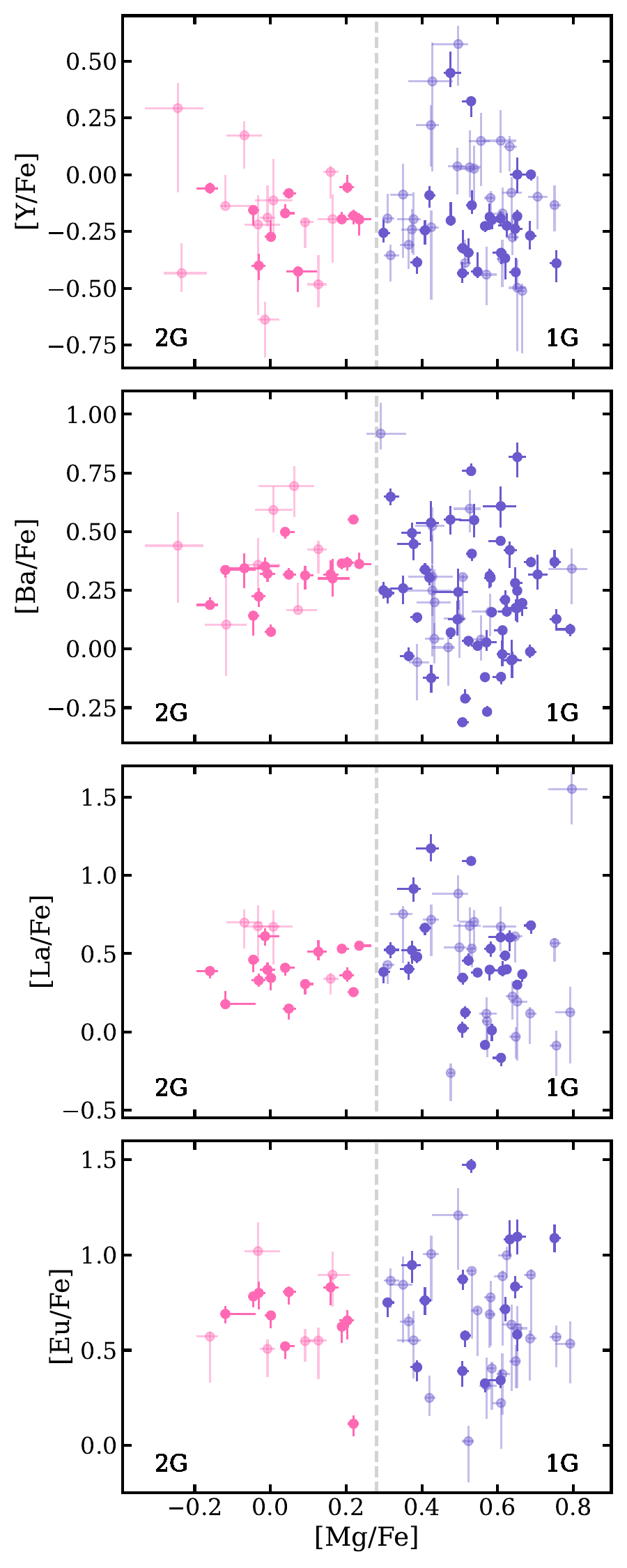}
    \caption{Neutron-capture abundances in M15 with error $<$ 0.4 dex. The data are colored by population with the split between populations 1 and 2 at $\rm{[Mg/Fe]}=0.28$, marked by the gray dashed line. Measurements with either upper or lower asymmetric error $>$ 0.1 are less opaque.}
    \label{fig:XFevMgFe}
\end{figure}

We defined the first generation (1G) and second generation (2G) of stars in M15 based on [Mg/Fe]. The Na and Al measurements for these stars were not of good quality; the Na D lines (5890 and 5896 \AA\@) were excluded from the model to avoid interstellar contamination, and Al is known to be strongly affected by NLTE in metal-poor stars \citep{gehren_NLTE_2004, andrievsky_AlNLTE_2008}. C and N do not have measurable atomic lines within the wavelength range and resolution of our spectra, and the CN feature around 4215 \AA\@ is not sensitive enough to N to use the C-N anticorrelation. Without other light element abundances, the light element correlations could not be used to define the populations. Instead, we used k-means clustering to identify the two populations based on [Mg/Fe], implemented with \texttt{scikit-learn} \citep{macqueen_kmeans_1967, scikit-learn}. The results of this are shown in Figure \ref{fig:Kmeans}, with the value of $\rm{[Mg/Fe]}=0.28$ that separates the two populations marked. The abundances in Figure \ref{fig:XFevMgFe} are colored by population with the split at $\rm{[Mg/Fe]}=0.28$.

We used MCMC to estimate the mean and standard deviation of the abundance of each element within each generation assuming that the abundances are normally distributed \citep{emcee_2013}. We implemented two different error cuts on the abundances: one at 0.1 dex and the other at 0.4 dex. The results of these calculations for each neutron-capture element considered are presented in Table \ref{tab:mcmcresults}.

In Figure \ref{fig:XFevMgFe}, it is clear that there is an outlier with low [Eu/Fe] in 2G (star 54055). This star passed all membership cuts (see Section \ref{sec:data}) and the abundance measurements pass quality cuts. With its low Eu abundance, perhaps it is a member of 1G with low Mg. We performed the MCMC mean and standard deviation calculations for Eu twice for each error cut, once with and once without 54055. This is noted in Table \ref{tab:mcmcresults}.  

The Y abundance dispersions in 1G and 2G are not significantly different. In contrast, the Ba and La dispersions in 1G and 2G are different at the 2 $\sigma$ level for both error cuts. The same is true of Eu when the star 54055 is left out. When 54055 is included, the calculated standard deviation of the 2G abundances increases and is consistent with that of 1G within 1 $\sigma$.

\section{Discussion} \label{sec:discussion}

\subsection{Implications for multiple populations} \label{subsec:MPimplications}

The member-to-member variations in light element abundances (i.e., multiple populations in GCs) may have already been present in the interstellar gas at the time of star formation, or they may have arisen at a later time from the exchange of material between already formed stars. The former scenario necessarily requires multiple episodes of star formation in the GC, as is typically assumed in \textit{multiple generation models} (e.g., \citealt{early_NaAl_1,FRMS,SMS}). In the latter case, only one generation of stars is required, as in \textit{concurrent formation models} (e.g., \citealt{early_disk_accretion,concurrent_SMS,BD_accretion}). It is also possible that both scenarios contribute to some extent, and that relative contributions vary among GCs.

Observations of $r$-process dispersion in GCs support the multiple generation scenario. As in M92 and NGC 2298 \citep{kirby_r-process_2023,bandyopadhyay_ngc2298_2025}, the generation-dependent abundance dispersions in M15 suggest that the two generations of stars formed at different times, with enough time between the formations of 1G and 2G for the gas to sufficiently mix. This also shows that the gas that produced the 1G stars was not well-mixed when the stars were forming.

Our finding is consistent with earlier photometric studies of GCs that showed that the range of light element abundances in GCs is largely independent of stellar mass \citep{gerasimov_47Tuc_2024,no_mass_variations_1,no_mass_variations_3,no_mass_variations_2}, which suggests that the gas of later stellar generations was enriched prior to star formation (i.e., before stellar masses were ``set''). However, deeper photometry with the James Webb Space Telescope has revealed that lower-Main Sequence stars near the hydrogen-burning limit may have a distinct distribution of light element abundances from their higher-mass counterparts in the GCs M4, NGC 6397 and 47\,Tuc \citep{2024A&A...690A.371L,2025arXiv250713564B,2025A&A...694A..68S}. 47\,Tuc in particular exhibits a prominent discontinuity in the lower Main Sequence \citep{2024ApJ...965..189M,2025A&A...694A..68S}, which most likely indicates that the lowest-mass stars in the cluster are overwhelmingly oxygen-poor (2G), allowing rapid formation of methane in the stellar atmosphere at lower temperatures. While this observation may be partially explained by dynamical effects (e.g., preferential mass segregation and evaporation of kinematically distinct low-mass 1G stars), the magnitude of the effect suggests that concurrent formation models cannot be completely ruled out. This is particularly important in the low-mass regime ($\lesssim 0.1 M_{\sun}$), in which the accretion rate of young pre-Main Sequence stars is expected to be orders of magnitude slower than that of higher-mass cluster members \citep{accretion_rate_1,accretion_rate_2}. Slower accretion may allow low-mass 1G stars to be polluted by oxygen-poor material produced by fast sources of chemical enrichment such as fast-rotating massive stars \citep{FRMS,chiappini_FRMS_2011}, massive binary stars \citep{massive_binaries}, or supermassive stars \citep{SMS}.

M15 is now one of three metal-poor GCs that have been shown to have a higher $r$-process dispersion in 1G than in 2G\@. This pattern is therefore not unique to any one cluster but appears to be fairly common among clusters---especially very metal-poor clusters---that display neutron-capture abundance dispersions in the first place. It is likely that a complete description of the origin of multiple populations in GCs involves a combination of multiple generation and concurrent formation approaches, as well as dynamical effects.

\subsection{Implications for the $r$-process} \label{subsec:rprocimplications}
If the $r$-process event that enriched M15 was from the first generation as some of the stars were still forming, the event had to have a short delay time. Potential sites of the $r$-process with a short delay time include magnetorotational supernovae \citep{nishimura_r-process_2015}, collapsar jets \citep{mumpower_jets_2025}, and other exotic types of core-collapse supernovae.

Another possible interpretation of the different $r$-process dispersions between generations is that the progenitor gas cloud of M15 had already been polluted by an $r$-process event and did not have enough time to mix before the 1G stars formed. In this case, the constraints on the delay time of the $r$-process event are not as stringent.

M15, M92, and NGC~2298 all exhibit a smaller dispersion for the first-peak (``limited'' $r$-process, such as Sr, Y, and Zr) elements compared to Ba and the lanthanides (``main'' $r$-process, such as Ba, La, and Eu). This different behavior could indicate that the main $r$-process is generated in fewer, rarer sources than the limited $r$-process, which might be more commonly produced, perhaps even in ordinary core collapse supernovae \citep[e.g., the $\nu p$-process,][]{frohlich_nup_2006}.

\vspace{2em}
We are grateful to L.\ Arielle Phillips for her encouragement and direction during the beginning stages of this project.

The data presented herein were obtained at Keck Observatory, which is a private 501(c)3 non-profit organization operated as a scientific partnership among the California Institute of Technology, the University of California, and the National Aeronautics and Space Administration. The Observatory was made possible by the generous financial support of the W. M. Keck Foundation.  The authors wish to recognize and acknowledge the very significant cultural role and reverence that the summit of Maunakea has always had within the Native Hawaiian community. We are most fortunate to have the opportunity to conduct observations from this mountain.

\vspace{5mm}

\facility{Keck:II (DEIMOS)}

\software{\texttt{astropy} \citep{astropy_I, astropy_II, astropy_III},
\texttt{emcee} \citep{emcee_2013},
\texttt{matplotlib} \citep{Hunter_matplotlib_2007},
\texttt{scikit-learn} \citep{scikit-learn},
\texttt{scipy} \citep{scipy}, 
\texttt{spec2d} pipeline \citep{cooper_spec2d_2012,newman_deep2_2013},
\texttt{ATLAS-9} \citep{ATLAS5,ATLAS9_1,ATLAS9_2},
\texttt{SYNTHE} \citep{SYNTHE},
\texttt{BasicATLAS} \citep{BasicATLAS}
    }

\bibliography{SMAUG_refs_ADS}{}
\bibliographystyle{aasjournal}

\end{document}